\documentclass[%
 aapm,footnoteinbib,
 mph,%
 amsmath,amssymb,longbibliography,reprint%
]{revtex4-1}

\usepackage{graphicx}% Include figure files
\usepackage{dcolumn}% Align table columns on decimal point
\usepackage{bm}% bold math
\usepackage{amsmath,amssymb}

\usepackage{hyperref}
\usepackage[]{natbib}

\usepackage{tikz}
\usepackage{lipsum}

\begin{document}

\title{Two concurrent exponential decays in a vacuum capacitor-resistor circuit: effect of surface charge}

%ID for IEEE conference is cscc2024-238 


\author{Frank V. Kowalski}%
\email{fkowalsk@mines.edu}

\affiliation{Physics Department, Colorado
School of Mines, Golden CO. 80401 U.S.A.}

\begin{abstract}
The relaxation in a vacuum capacitor-resistor circuit is comprised of two exponential decays, one caused by surface charge and the other by the decay of energy stored between the capacitor plates. A simple phenomenological model of this relaxation is shown to be supported by measurements even though Maxwell's equations are difficult to apply in this case. Similar behavior is also observed for polypropylene capacitors, indicating that this surface charge effect is applicable to all capacitors and potentially other circuit components.
\end{abstract}

\date{\today}% It is always \today, today,

 
 \maketitle

\section{\label{sec:level1} Introduction}

%add sections, add frequency domain graphs, emphasize effect of charges on surface of circuit components affecting behavior

The response of a capacitor is modeled with Kirchhoff's laws. These predict the ``ideal'' behavior associated with equal and opposite charges on the capacitor plates and exponential relaxation in an RC circuit. The same current flows into one terminal of the capacitor and out of the other.

However, non-exponential decay occurs when the capacitor contains a dielectric.\cite{kohlrausch} Such a response is used to understand the structure of the dielectric. \cite{kramer,jonscher,feldman} This continues to be a topic of interest. \cite{westerlund,2022,2023}

This effect is not present when using a vacuum capacitor. Nevertheless, a mechanism that does lead to non-ideal behavior is the equivalent series inductance of such a device that typically produces self-resonance frequencies above $100$ MHz. \cite{murata} The results presented here occur at much lower frequencies. 

A first principles model of circuits is provided by Maxwell's equations. The few such solutions found in the literature involve steady state behavior. \cite{muller} The circuit wires must then have a surface charge to constrain current flow within the wires. \cite{sommerfeld, heald, chabay,moreau}  This surface charge has been shown to behave as a Hall probe. \cite{schade} A numerical solution to Maxwell's equations for decay in an RC circuit does not account for the results presented below. \cite{preyer}

An example of the effect of surface charge involves a knee in a circuit wire. Assume greater current into the knee than out of it,  ``then the charge piles up at the ``knee,'' and this produces a field aiming away at the knee. The field opposes the current flowing in (slows it down) and promotes the current flowing out (speeding it up) until these currents are equal, at which point there is no further accumulation of charge and equilibrium is established.'' \cite{griffiths2} The data below illustrate a similar effect where the current flowing into the capacitor differs from that flowing out.  

It is difficult to obtain first principles solutions for circuits that exhibit transient behavior. \cite{klee,muller,heald} Nevertheless, a phenomenological model for the observed two concurrent exponential decays in a vacuum capacitor-resistor circuit is presented, one decay of which is shown to be associated with surface charge.

\section{\label{sec:level2} Results}

\begin{figure}[ht]
{%
  \includegraphics[width=.97 \columnwidth]{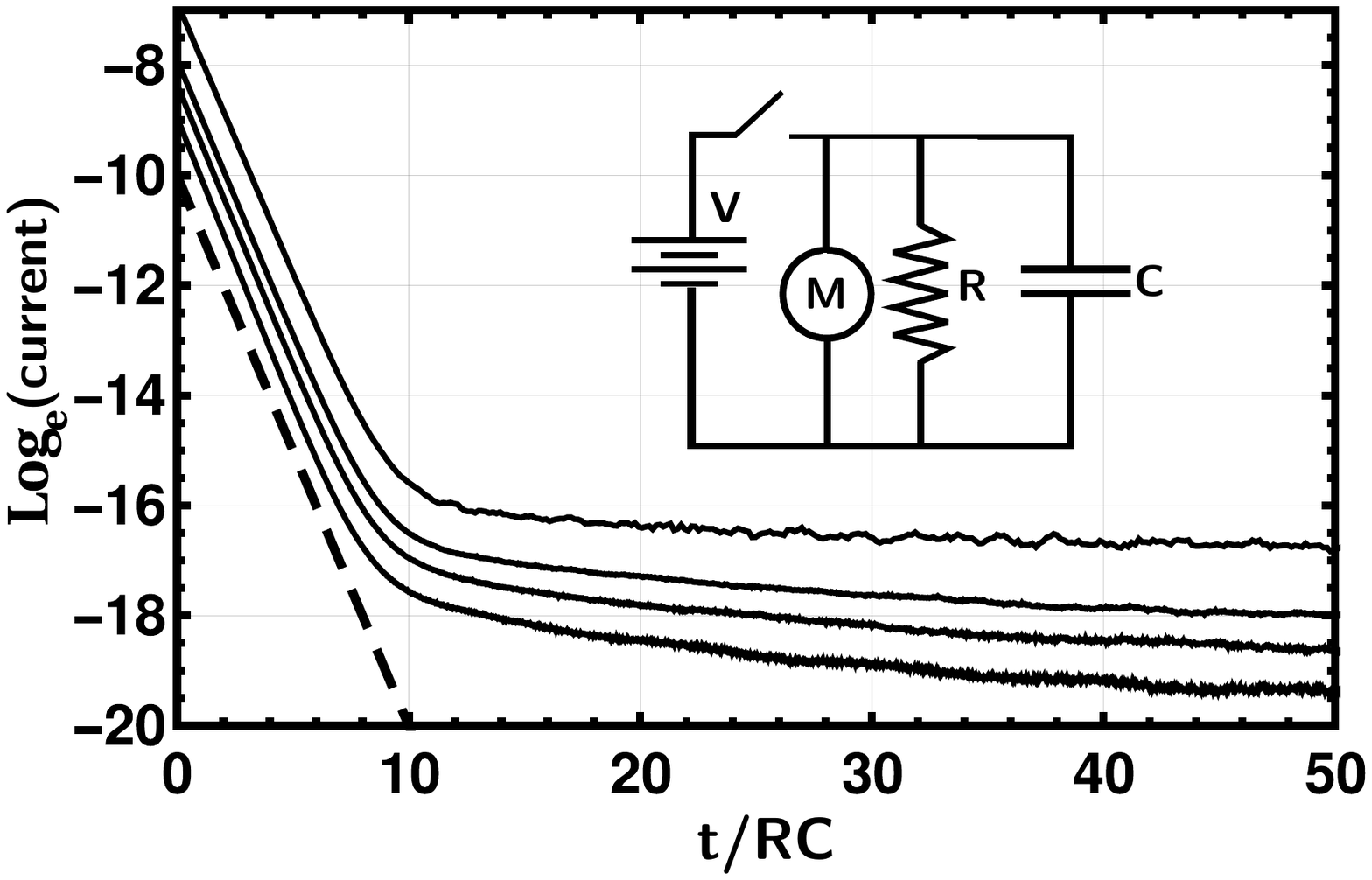}%
}\hfill
\caption{Logarithm of the current vs time in time constant units for the circuit in the inset. The lines from top to bottom correspond to data for increasing resistances while the capacitance is fixed. The expected RC decay for the lowest trace is shown as the dashed line. It is offset for clarity. }
\label{fig1}
\end{figure}

Fig \ref{fig1} shows current decays for fixed capacitance with different resistance values for the circuit in the inset: $C=2.789$ nF (Comet model CFMN-2800BAC/8-DE-G) while resistances from top to bottom are $R=47~\mathrm{k}\Omega$, $R=182~\mathrm{k}\Omega$, $R=337~\mathrm{k}\Omega$,  and $R=637~\mathrm{k}\Omega$. These data deviate from the expected single exponential decay that is shown as the dashed line. A $98$ V power supply energizing the circuit was disconnected using a mechanical switch. A Keysight 34465A digital voltmeter was used to collect all the data presented below.

\begin{figure}[!]
{%
  \includegraphics[width=.97 \columnwidth]{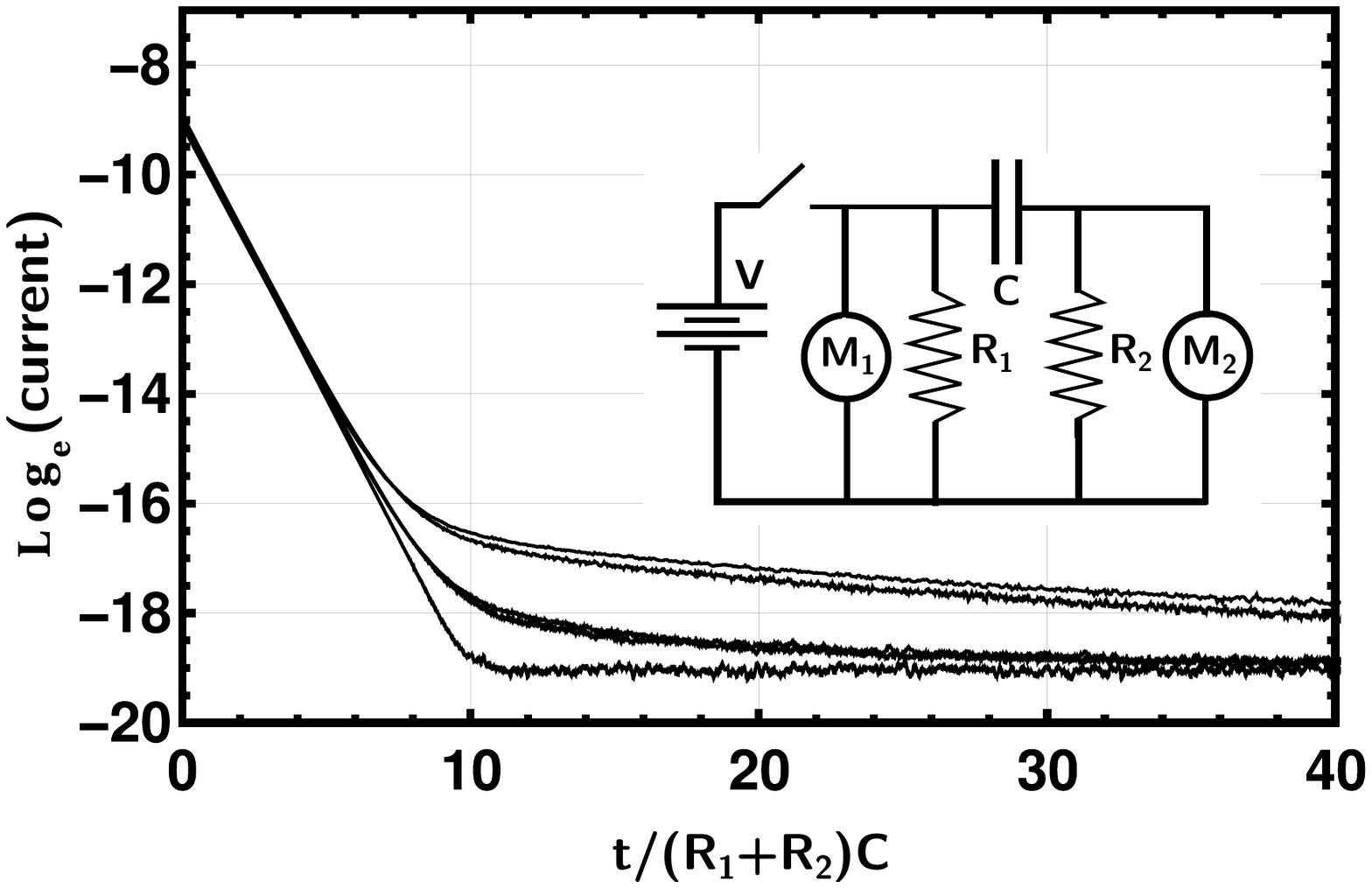}%
}\hfill
\caption{Current relaxation for the circuit in the inset for fixed resistance $R_{1}=R_{2}=337.8~{\textrm k}\Omega$ while C varies. The trace just above the bottom trace consists of two decays that overlap. The bottom trace corresponds to currents through $R_{2}$ for all four values of capacitance while the other traces are for the currents through $R_{1}$. The data are from reference \cite{kowalski}.}
\label{fig6}
\end{figure}

Fig. \ref{fig6} illustrates the decay for fixed $R_{1}=R_{2}=337.8~{\textrm k}\Omega$ while C varies. These data are obtained using four vacuum capacitors, two with C$=0.5177$ nF (model CFMN-500AAC/12-DE-G) and two with C$=2.789$ nF (Comet model CFMN-2800BAC/8-DE-G). The upper trace is the current through $R_{1}$ with C$=0.5177$ nF and the trace just below it (that almost overlaps with the upper trace) is the current through $R_{1}$ for C$=1.0354$ nF (a parallel combination of two C$=0.5177$ nF capacitors). The next two traces below these are for the current through $R_{1}$ with C$=2.789$ nF and C$=5.578$ nF (a parallel combination of two C$=2.789$ nF capacitors). These traces completely overlap. The lowest trace is the decay data for the current through $R_{2}$ for all four values of C.

The unequal currents into and out of the capacitor shown in fig. \ref{fig6} result in a net charge on the capacitor. However, the initial charge on each capacitor plate is not measured.

\section{Determination of variables}

Decay data expressed in terms of the dimensionless variable $\tau=$ t/RC facilitates the determination of the variables that influence the decay. For example, if the relaxation is a function of R and C then the current at a particular value of $\tau$ will be the same for different combinations of R and C that yield this value of $\tau$. All plots of the decay versus time for different values of R and C coalesce into only one when plotted versus $\tau$. 

This, however, is not the case if the decay depends on another variable. R and C are the sole variables influencing the decay for the faster but not slower decay shown in fig. \ref{fig1} (apart from the offset due to different initial currents). However, a model that the slower decay data supports must include more variables than R and C. 

Consider this argument for fixed resistance while the capacitance varies as shown in fig. \ref{fig6} (in this case the initial current is the same for all variations of capacitance leading to overlap of all the faster decay data). The upper pair of the traces (for C$=0.5177$ nF and C$=1.0354$ nF) are assumed (for the sake of argument) to overlap. The doubling of C (by adding an identical capacitor in parallel) does not change the plot (it does in a plot of current versus time). This implies that R and C are the only variables required to model the data (assuming for the moment that variation in R also yields the same plot). 

%This model, a function of $\tau=$ t/RC, is not that of single exponential decay.

Similarly, the overlap of the plots of the current decay vs $\tau$ for C$=2.789$ nF and C$=5.578$ nF implies that R and C are the only variables required to model the decay from these capacitors (again assuming for the moment that variation in R also yields the same plot). However, a new variable must exist to model both pairs of data since the decay differs for the two pairs of lines. 

One then looks for a variable that is the same for each pair but differs between them. Each pair of lines is generated first by one capacitor and second by the same capacitor in parallel with another identical capacitor. The geometry of the capacitors used to generate one pair of lines differs from that of the capacitors used to generated the other pair of lines and is therefore such a variable.

\section{\label{sec:level3} Phenomenological Model}

A model of the current from one capacitor plate is given by $I=I_{10} \exp[-\alpha t]+I_{20} \exp[-\beta t]$. This is referred to as the sum of exponentials model. The parameter $\alpha=1/\textrm{RC}$ is determined from measurements of R and C where R is the total resistance in the circuit. The parameters $I_{20}$ and $\beta$ are determined for each capacitor plate from a fit of the current decay data to this model. Such fits can be found for both vacuum and polypropylene capacitors in reference \cite{kowalski}.

\section{\label{sec:level5} Surface Charge Model}

A model based on a microscopic understanding of this decay separates the amount of charge on a capacitor plate into two parts: one associated with the energy stored in between the capacitor plates, $Q^{in}$, whose electric field is confined in between the plates, and $Q^{out}$, the charge on the circuit wire and capacitor plate that facilitates the flow of current in the circuit with its electric field not confined. The superposition principle yields a response that is a sum of responses for each type of charge in this separate charge model. Consider the voltage expression for the sum of exponentials model. Let the $V_{10} \exp[-\alpha t]$ term be associated with $Q^{in}$.  Let $Q^{out}_{0}$ be the charge required to direct the current from the power supply into $R$ before the switch is opened. 

One might conjecture that this takes the form $Q^{out}_{0}=V^{out}_{0} \tau_{out}/ R$, where $\tau_{out}$ is a constant and $V^{out}_{0}$ is the voltage at the plate generated by $Q^{out}_{0}$. For smaller $R$ larger $Q^{out}_{0}$ is required to direct this larger current away from the capacitor plate and into the resistor (similar to the effect that charge on the hairpin wire has on the current flowing through the wire). 

After the switch opens $V^{out}=R dQ^{out}/dt$. The loss of $Q^{out}$ is then determined from $dQ^{out}/dt=-Q^{out}/\tau_{out}$ leading to exponential decay. The parameters in the model term $V_{20} \exp[-\beta t]$ are $V_{20}=V^{out}_{0}$ and $\beta=1/\tau_{out}$. 

For the circuit in fig. \ref{fig6} the left hand capacitor plates have non-zero $Q^{out}_{0}$. Surface charge is needed on this side of the capacitor to redirect the current from the power supply through the resistor before the switch is opened. In comparison, all the right hand capacitor plates have essentially no $Q^{out}$ (since there is no such initial current to redirect) as indicated by the lack of a slow decay from these plates. 

However, $Q^{in}$ and $\alpha$ are the same for both plates of a particular capacitor resulting in the same fast exponential decays of $Q^{in}$ for both the left and right hand plates. This is illustrated in fig. \ref{fig6} for all capacitor values. 

Different capacitors yield different $\beta$ values as shown in fig. \ref{fig6}. This dependence of the slow decay on the capacitor geometry is consistent with the conclusion from the above dimensionless variable analysis. The sum of charges model then provides a microscopic justification for the sum of exponentials model and for the dimensional analysis argument about the variables influencing the relaxation.

\section{\label{sec:level6} Discussion}

The geometry of a circuit increases the difficulty in calculating the surface charge from Maxwell's equations. \cite{muller} Even constructing a numerical model that predicts only a single exponential decay for an RC circuit from Maxwell's equations is nontrivial. \cite{preyer} Although the data illustrate similar but simple (a sum of exponentials) behavior for geometrically dissimilar capacitors it is difficult to determine $Q^{out}$ from first principles. In addition, the sum of exponentials model does not indicate a coupling between the decays of $Q^{in}$ and $Q^{out}$.

More effort is required to better understand the effect of surface charge on circuits. For example, the dependence of $\beta$ on the energizing voltage has not been addressed. It seems likely that $Q^{out}$ also influences the behavior of inductors. Also of interest is how surface charge affects circuits which decay radiatively (rather than thermally) in for example a superconducting LC circuit. The effect of surface charge on the conversion of electrical energy into mechanical energy (or into more complicated circuit components) is also of interest.

\begin{acknowledgments}
I wish to thank Justin L. Swantek, Tony D'Esposito, and Jacob Brannum for useful discussions. The support from a HP Technology for Teaching Grant is acknowledged.
\end{acknowledgments}

%\section*{Data Availability}
%The data that support the findings of this study are available from the corresponding author upon reasonable request.

%\section*{Conflict of interest}
%The author has no conflicts to disclose.

 \hspace{20 mm}

\clearpage

%\nocite{*}
%\bibliography{aapmsamp}

\begin{thebibliography}{23}%
\makeatletter
\providecommand \@ifxundefined [1]{%
 \@ifx{#1\undefined}
}%
\providecommand \@ifnum [1]{%
 \ifnum #1\expandafter \@firstoftwo
 \else \expandafter \@secondoftwo
 \fi
}%
\providecommand \@ifx [1]{%
 \ifx #1\expandafter \@firstoftwo
 \else \expandafter \@secondoftwo
 \fi
}%
\providecommand \natexlab [1]{#1}%
\providecommand \enquote  [1]{``#1''}%
\providecommand \bibnamefont  [1]{#1}%
\providecommand \bibfnamefont [1]{#1}%
\providecommand \citenamefont [1]{#1}%
\providecommand \href@noop [0]{\@secondoftwo}%
\providecommand \href [0]{\begingroup \@sanitize@url \@href}%
\providecommand \@href[1]{\@@startlink{#1}\@@href}%
\providecommand \@@href[1]{\endgroup#1\@@endlink}%
\providecommand \@sanitize@url [0]{\catcode `\\12\catcode `\$12\catcode
  `\&12\catcode `\#12\catcode `\^12\catcode `\_12\catcode `\%12\relax}%
\providecommand \@@startlink[1]{}%
\providecommand \@@endlink[0]{}%
\providecommand \url  [0]{\begingroup\@sanitize@url \@url }%
\providecommand \@url [1]{\endgroup\@href {#1}{\urlprefix }}%
\providecommand \urlprefix  [0]{URL }%
\providecommand \Eprint [0]{\href }%
\providecommand \doibase [0]{http://dx.doi.org/}%
\providecommand \selectlanguage [0]{\@gobble}%
\providecommand \bibinfo  [0]{\@secondoftwo}%
\providecommand \bibfield  [0]{\@secondoftwo}%
\providecommand \translation [1]{[#1]}%
\providecommand \BibitemOpen [0]{}%
\providecommand \bibitemStop [0]{}%
\providecommand \bibitemNoStop [0]{.\EOS\space}%
\providecommand \EOS [0]{\spacefactor3000\relax}%
\providecommand \BibitemShut  [1]{\csname bibitem#1\endcsname}%
\let\auto@bib@innerbib\@empty
%</preamble>
\bibitem [{\citenamefont {Kohlrausch}(2009)}]{kohlrausch}%
  \BibitemOpen
  \bibfield  {author} {\bibinfo {author} {\bibfnamefont {R.}~\bibnamefont
  {Kohlrausch}},\ }\bibfield  {title} {\enquote {\bibinfo {title} {Theorie des
  elektrischen r{\"u}ckstandes in der leidener flasche},}\ }\href
  {https://doi.org/10.1002/andp.18541670103} {\bibfield  {journal} {\bibinfo
  {journal} {Ann. Phys.}\ }\textbf {\bibinfo {volume} {167}},\ \bibinfo {pages}
  {56--58} (\bibinfo {year} {2009}).}\BibitemShut {NoStop}%
\bibitem [{\citenamefont {Kramer}(2012)}]{kramer}%
  \BibitemOpen
  \bibfield  {author} {\bibinfo {author} {\bibfnamefont {F.}~\bibnamefont
  {Kramer}},\ }\href@noop {} {\emph {\bibinfo {title} {Broadband Dielectric
  Spectroscopy}}}\ (\bibinfo  {publisher} {Springer Berlin Heidelberg},\
  \bibinfo {year} {2012})\ pp.\ \bibinfo {pages} {48--51}\BibitemShut {NoStop}%
\bibitem [{\citenamefont {Jonscher}(1999)}]{jonscher}%
  \BibitemOpen
  \bibfield  {author} {\bibinfo {author} {\bibfnamefont {A.~K.}\ \bibnamefont
  {Jonscher}},\ }\bibfield  {title} {\enquote {\bibinfo {title} {Dielectric
  relaxation in solids},}\ }\href {\doibase 10.1088/0022-3727/32/14/201}
  {\bibfield  {journal} {\bibinfo  {journal} {Journal of Physics D: Applied
  Physics}\ }\textbf {\bibinfo {volume} {32}},\ \bibinfo {pages} {R57--R70}
  (\bibinfo {year} {1999})}\BibitemShut {NoStop}%
\bibitem [{\citenamefont {Feldman}\ \emph {et~al.}(2005)\citenamefont
  {Feldman}, \citenamefont {Puzenko},\ and\ \citenamefont {Ryabov}}]{feldman}%
  \BibitemOpen
  \bibfield  {author} {\bibinfo {author} {\bibfnamefont {Y.}~\bibnamefont
  {Feldman}}, \bibinfo {author} {\bibfnamefont {A.}~\bibnamefont {Puzenko}}, \
  and\ \bibinfo {author} {\bibfnamefont {Y.}~\bibnamefont {Ryabov}},\
  }\bibfield  {title} {\enquote {\bibinfo {title} {Dielectric relaxation
  phenomena in complex materials},}\ }\href {\doibase 10.1002/0471790265.ch1}
  {\bibfield  {journal} {\bibinfo  {journal} {Advances in Chemical Physics}\
  }\textbf {\bibinfo {volume} {133}},\ \bibinfo {pages} {1 -- 125} (\bibinfo
  {year} {2005})}\BibitemShut {NoStop}%
\bibitem [{\citenamefont {Westerlund}\ and\ \citenamefont
  {Ekstam}(1994)}]{westerlund}%
  \BibitemOpen
  \bibfield  {author} {\bibinfo {author} {\bibfnamefont {S.}~\bibnamefont
  {Westerlund}}\ and\ \bibinfo {author} {\bibfnamefont {L.}~\bibnamefont
  {Ekstam}},\ }\bibfield  {title} {\enquote {\bibinfo {title} {Capacitor
  theory},}\ }\href {\doibase 10.1109/94.326654} {\bibfield  {journal}
  {\bibinfo  {journal} {IEEE Transactions on Dielectrics and Electrical
  Insulation}\ }\textbf {\bibinfo {volume} {1}},\ \bibinfo {pages} {826--839}
  (\bibinfo {year} {1994})}\BibitemShut {NoStop}%
\bibitem [{\citenamefont {{Allagui}}\ \emph {et~al.}(2022)\citenamefont
  {{Allagui}}, \citenamefont {{Zhang}},\ and\ \citenamefont
  {{Elwakil}}}]{2022}%
  \BibitemOpen
  \bibfield  {author} {\bibinfo {author} {\bibfnamefont {Anis}\ \bibnamefont
  {{Allagui}}}, \bibinfo {author} {\bibfnamefont {Di}~\bibnamefont {{Zhang}}},
  \ and\ \bibinfo {author} {\bibfnamefont {Ahmed}\ \bibnamefont {{Elwakil}}},\
  }\bibfield  {title} {\enquote {\bibinfo {title} {{Further Experimental
  Evidence of the Dead Matter Has Memory Conjecture in Capacitive Devices}},}\
  }\href {\doibase 10.48550/arXiv.2206.15043} {\bibfield  {journal} {\bibinfo
  {journal} {arXiv e-prints}\ ,\ \bibinfo {eid} {arXiv:2206.15043}} (\bibinfo
  {year} {2022})},\ \Eprint {http://arxiv.org/abs/2206.15043} {arXiv:2206.15043
  [physics.app-ph]} \BibitemShut {NoStop}%
\bibitem [{\citenamefont {Ortigueira}\ \emph {et~al.}(2023)\citenamefont
  {Ortigueira}, \citenamefont {Martynyuk}, \citenamefont {Kosenkov},\ and\
  \citenamefont {Batista}}]{2023}%
  \BibitemOpen
  \bibfield  {author} {\bibinfo {author} {\bibfnamefont {Manuel}\ \bibnamefont
  {Ortigueira}}, \bibinfo {author} {\bibfnamefont {Valeriy}\ \bibnamefont
  {Martynyuk}}, \bibinfo {author} {\bibfnamefont {Volodymyr}\ \bibnamefont
  {Kosenkov}}, \ and\ \bibinfo {author} {\bibfnamefont {Arnaldo}\ \bibnamefont
  {Batista}},\ }\bibfield  {title} {\enquote {\bibinfo {title} {A new look at
  the capacitor theory},}\ }\href {\doibase 10.3390/fractalfract7010086}
  {\bibfield  {journal} {\bibinfo  {journal} {Fractal and Fractional}\ }\textbf
  {\bibinfo {volume} {7}} (\bibinfo {year} {2023}),\
  10.3390/fractalfract7010086}\BibitemShut {NoStop}%
\bibitem [{\citenamefont {solutions}(2019)}]{murata}%
  \BibitemOpen
  \bibfield  {author} {\bibinfo {author} {\bibfnamefont {Cadence~PCB}\
  \bibnamefont {solutions}},\ }\href
  {https://resources.pcb.cadence.com/blog/2019-capacitor-self-resonant-frequency-and-signal-integrity}
  {\enquote {\bibinfo {title} {Capacitor self-resonant frequency and signal
  integrity},}\ } (\bibinfo {year} {2019})\BibitemShut {NoStop}%
\bibitem [{\citenamefont {Muller}(2012)}]{muller}%
  \BibitemOpen
  \bibfield  {author} {\bibinfo {author} {\bibfnamefont {R.}~\bibnamefont
  {Muller}},\ }\bibfield  {title} {\enquote {\bibinfo {title} {A
  semiquantitative treatment of surface charges in dc circuits},}\ }\href
  {\doibase 10.1119/1.4731722} {\bibfield  {journal} {\bibinfo  {journal}
  {American Journal of Physics}\ }\textbf {\bibinfo {volume} {80}},\ \bibinfo
  {pages} {782--788} (\bibinfo {year} {2012})},\ \Eprint
  {http://arxiv.org/abs/https://doi.org/10.1119/1.4731722}
  {https://doi.org/10.1119/1.4731722} \BibitemShut {NoStop}%
\bibitem [{\citenamefont {Sommerfeld}(1952)}]{sommerfeld}%
  \BibitemOpen
  \bibfield  {author} {\bibinfo {author} {\bibfnamefont {A.}~\bibnamefont
  {Sommerfeld}},\ }in\ \href@noop {} {\emph {\bibinfo {booktitle}
  {Electrodynamics}}}\ (\bibinfo  {publisher} {Academic, New York},\ \bibinfo
  {year} {1952})\ pp.\ \bibinfo {pages} {125--130}\BibitemShut {NoStop}%
\bibitem [{\citenamefont {Heald}(1984)}]{heald}%
  \BibitemOpen
  \bibfield  {author} {\bibinfo {author} {\bibfnamefont {M.~A.}\ \bibnamefont
  {Heald}},\ }\bibfield  {title} {\enquote {\bibinfo {title} {Electric fields
  and charges in elementary circuits},}\ }\href {\doibase 10.1119/1.13611}
  {\bibfield  {journal} {\bibinfo  {journal} {American Journal of Physics}\
  }\textbf {\bibinfo {volume} {52}},\ \bibinfo {pages} {522--526} (\bibinfo
  {year} {1984})},\ \Eprint
  {http://arxiv.org/abs/https://doi.org/10.1119/1.13611}
  {https://doi.org/10.1119/1.13611} \BibitemShut {NoStop}%
\bibitem [{\citenamefont {Chabay}\ and\ \citenamefont
  {Sherwood}(2019)}]{chabay}%
  \BibitemOpen
  \bibfield  {author} {\bibinfo {author} {\bibfnamefont {R.}~\bibnamefont
  {Chabay}}\ and\ \bibinfo {author} {\bibfnamefont {B.}~\bibnamefont
  {Sherwood}},\ }\bibfield  {title} {\enquote {\bibinfo {title} {Polarization
  in electrostatics and circuits: Computing and visualizing surface charge
  distributions},}\ }\href {\doibase 10.1119/1.5095939} {\bibfield  {journal}
  {\bibinfo  {journal} {American Journal of Physics}\ }\textbf {\bibinfo
  {volume} {87}},\ \bibinfo {pages} {341--349} (\bibinfo {year} {2019})},\
  \Eprint {http://arxiv.org/abs/https://doi.org/10.1119/1.5095939}
  {https://doi.org/10.1119/1.5095939} \BibitemShut {NoStop}%
\bibitem [{\citenamefont {Moreau}(1989)}]{moreau}%
  \BibitemOpen
  \bibfield  {author} {\bibinfo {author} {\bibfnamefont {W~R}\ \bibnamefont
  {Moreau}},\ }\bibfield  {title} {\enquote {\bibinfo {title} {Charge
  distributions on dc circuits and kirchhoff's laws},}\ }\href {\doibase
  10.1088/0143-0807/10/4/008} {\bibfield  {journal} {\bibinfo  {journal}
  {European Journal of Physics}\ }\textbf {\bibinfo {volume} {10}},\ \bibinfo
  {pages} {286} (\bibinfo {year} {1989})}\BibitemShut {NoStop}%
  \bibitem [{\citenamefont {Schade}\ \emph {et~al.}(2019)\citenamefont {Schade},
  \citenamefont {Schuster},\ and\ \citenamefont {Nagel}}]{schade}%
  \BibitemOpen
  \bibfield  {author} {\bibinfo {author} {\bibfnamefont {Nicholas}\
  \bibnamefont {Schade}}, \bibinfo {author} {\bibfnamefont {David}\
  \bibnamefont {Schuster}}, \ and\ \bibinfo {author} {\bibfnamefont {Sidney}\
  \bibnamefont {Nagel}},\ }\bibfield  {title} {\enquote {\bibinfo {title} {A
  nonlinear, geometric hall effect without magnetic field},}\ }\href {\doibase
  10.1073/pnas.1916406116} {\bibfield  {journal} {\bibinfo  {journal}
  {Proceedings of the National Academy of Sciences}\ }\textbf {\bibinfo
  {volume} {116}},\ \bibinfo {pages} {201916406} (\bibinfo {year}
  {2019})}\BibitemShut {NoStop}%
\bibitem [{\citenamefont {Preyer}(2002)}]{preyer}%
  \BibitemOpen
  \bibfield  {author} {\bibinfo {author} {\bibfnamefont {N.~W.}\ \bibnamefont
  {Preyer}},\ }\bibfield  {title} {\enquote {\bibinfo {title} {Transient
  behavior of simple rc circuits},}\ }\href {\doibase 10.1119/1.1508444}
  {\bibfield  {journal} {\bibinfo  {journal} {American Journal of Physics}\
  }\textbf {\bibinfo {volume} {70}},\ \bibinfo {pages} {1187--1193} (\bibinfo
  {year} {2002})},\ \Eprint
  {http://arxiv.org/abs/https://doi.org/10.1119/1.1508444}
  {https://doi.org/10.1119/1.1508444} \BibitemShut {NoStop}%
\bibitem [{\citenamefont {Griffiths}(1989)}]{griffiths2}%
  \BibitemOpen
  \bibfield  {author} {\bibinfo {author} {\bibfnamefont {D.~J.}\ \bibnamefont
  {Griffiths}},\ }\href@noop {} {\emph {\bibinfo {title} {(Introduction to
  Electrodynamics: second edition).}}}\ (\bibinfo  {publisher}
  {Prentice-Hall},\ \bibinfo {year} {1989})\ pp.\ \bibinfo {pages}
  {277--278}\BibitemShut {NoStop}%
\bibitem [{\citenamefont {Klee}(2020)}]{klee}%
  \BibitemOpen
  \bibfield  {author} {\bibinfo {author} {\bibfnamefont {M.~M.}\ \bibnamefont
  {Klee}},\ }\bibfield  {title} {\enquote {\bibinfo {title} {Surface charges
  from a sensing pixel perspective},}\ }\href {\doibase 10.1119/10.0001435}
  {\bibfield  {journal} {\bibinfo  {journal} {American Journal of Physics}\
  }\textbf {\bibinfo {volume} {88}},\ \bibinfo {pages} {649--660} (\bibinfo
  {year} {2020})},\ \Eprint
  {http://arxiv.org/abs/https://doi.org/10.1119/10.0001435}
  {https://doi.org/10.1119/10.0001435} \BibitemShut {NoStop}%

  
  \bibitem{kowalski} From ``Energy relaxation in a vacuum capacitor-resistor circuit: measurement of multiple decays with divergent time constants'', by F. V. Kowalski, AIP Advances {\bf 14}, 035154 (2024) \url{https://doi.org/10.1063/5.0190681}. Licensed under a Creative Commons Attribution (CC BY) license (http://creativecommons.org/licenses/by/4.0/)  Adapted with permission.

  
  \end{thebibliography}
% Produces the bibliography via BibTeX.

\providecommand{\noopsort}[1]{}\providecommand{\singleletter}[1]{#1}%

\end{document}